\documentclass[prd,aps,nofootinbib,superscriptaddress,floatfix,preprintnumbers,twocolumn]{revtex4}
\usepackage[utf8]{inputenc}
\usepackage{amsfonts}
\usepackage{hyperref,amssymb,amsmath,graphicx,xcolor,slashed}
\usepackage[acronym]{glossaries}
\newacronym[longplural=parton distribution functions]{pdf}{PDF}{parton distribution function}

\newacronym{qcd}{QCD}{quantum chromodynamics}
\makeglossaries
\DeclareUnicodeCharacter{00A0}{ }

\newcommand{\PiP}[0]{(\bar{u}\gamma_5 d)}
\newcommand{\PiPmu}[0]{(\bar{u}\gamma_{\mu} \gamma_5 d)}
\newcommand{\PiPt}[0]{(\bar{u}\gamma_t \gamma_5 d)}
\newcommand{\PiAV}[0]{(\bar{u}\gamma_z \gamma_5 d)}
\newcommand{\PiPlus}[0]{(\bar{u}\gamma_+ \gamma_5 d)}

\usepackage[normalem]{ulem}
\begin{document}

\title{Kinematically-enhanced interpolating operators for boosted hadrons}

\author{Rui Zhang}
\email{rzhang93@mit.edu}
\affiliation{Physics Division, Argonne National Laboratory, Lemont, IL 60439, USA}

\author{Anthony V. Grebe}
\affiliation{Fermi National Accelerator Laboratory, Batavia, IL 60510, USA}  
 
\author{Daniel C. Hackett}
\affiliation{Fermi National Accelerator Laboratory, Batavia, IL 60510, USA} 

\author{Michael L. Wagman}
 \affiliation{Fermi National Accelerator Laboratory, Batavia, IL 60510, USA}  

\author{Yong Zhao}
\affiliation{Physics Division, Argonne National Laboratory, Lemont, IL 60439, USA}

\preprint{FERMILAB-PUB-24-0968-T}

\begin{abstract}
    We propose to use 
    interpolating operators for lattice quantum chromodynamics (QCD) calculations of highly-boosted pions and nucleons
    with kinematically-enhanced ground-state overlap factors at large momentum.
    Because this kinematic enhancement applies to the signal but not the variance of the correlation function, these interpolating operators can achieve better signal-to-noise ratios at large momentum.
    We perform proof-of-principle calculations with boosted pions and nucleons using close-to-physical and larger quark masses to explore the utility of our proposal.
    Results for effective energies and matrix elements, as well as Lanczos ground-state energy estimators, are consistent with theoretical expectations for signal-to-noise improvement at large momenta.
\end{abstract}
\maketitle
Boosted hadrons play a significant role in collider physics at the LHC~\cite{ATLAS:2019mfr} and the forthcoming Electron-Ion Collider (EIC)~\cite{Accardi:2012qut,AbdulKhalek:2021gbh}. Additionally, highly boosted light hadrons also frequently appear in other physical processes with large momentum transfer $Q^2$, such as the pion and kaon measurements at Jefferson Lab~\cite{Dudek:2012vr,jlab12GeV} and the future EIC~\cite{Aguilar:2019teb}, and in high-precision searches for unitary violations which relies on the decay of heavy mesons~\cite{BaBar:2001ags,Belle:2001zzw,Charles:2004jd,LHCb:2019hro} and baryons~\cite{LHCb:2025ray}.
Therefore, understanding the structure of boosted hadrons is crucial for advancing modern particle and nuclear physics.

{Lattice quantum chromodynamics (QCD) is essential for providing non-perturbative first-principles predictions for these experiments within the Standard Model~\cite{FlavourLatticeAveragingGroupFLAG:2024oxs}. To serve this goal, measurements of boosted hadrons on the lattice are necessary, especially in extracting their spin and three-dimension structures using near-lightcone approximations~\cite{Liu:1993cv,Detmold:2005gg,Braun:2007wv,Chambers:2017dov,Ji:2013dva,Ji:2013fga,Ji:2014gla,Yang:2016plb,Radyushkin:2017cyf,Ma:2017pxb,Detmold:2021uru,Ji:2020ect,Constantinou:2022yye,LatticePartonLPC:2023pdv}, and in measuring observables with large $Q^2$~\cite{Koponen:2017fvm,QCDSF:2017ssq,Davies:2018zav,FermilabLattice:2022gku,Parrott:2022rgu,Leskovec:2025gsw}.}
Recent lattice QCD calculations have employed pions boosted to $2.4$~GeV~\cite{Gao:2021dbh,Avkhadiev:2024mgd}, as well as kaons and nucleon boosted to $3.0$~GeV~\cite{Fan:2022kcb,LatticeParton:2022xsd,Ding:2024lfj}. 
However, a major computational challenge remains the reliable projection of hadronic states onto large momenta,  as the signal-to-noise ratios  (SNRs) for hadronic observables decrease rapidly at large boost, 
limiting the ability of lattice QCD calculations to reliably extract ground-state signals.
Techniques to improve the precision and reliability of lattice QCD calculations including highly boosted hadron states are thus extremely desirable.

Attempts to design interpolating operators (interpolators) that yield better signals for highly boosted hadron states have primarily focused on the spatial structure of the quark fields.
Early attempts included anisotropic spatial smearing resembling a ``plate''-like picture of hadrons in a Minkowski boosted frame~\cite{Roberts:2012tp,DellaMorte:2012xc}, but these operators did not show significant SNR improvement and seemed to worsen excited-state effects.
Significant progress was achieved through the proposal of momentum smearing quark propagators~\cite{Bali:2016lva}, which enhances both the SNR and overlap of standard hadron interpolators with boosted states. 
In this approach, the fermion fields are smeared with a phase factor to simulate a wave packet carrying specific momentum on the lattice.
Momentum smearing can lead to order-of-magnitude improvement in the SNR for large boosts, and its use is now standard in lattice QCD calculations of highly boosted hadrons~\cite{Egerer:2020hnc,Ji:2020ect,Constantinou:2022yye}.

Meanwhile, designing optimal spinor structures for highly boosted hadron interpolators has received less attention.
Standard interpolators acting on the vacuum create quark (antiquark) Fock states with the quantum numbers and symmetry properties of rest-frame hadrons and then multiply these states by a momentum phase factor.
Physically, highly boosted hadrons can be described by a lightcone field theory picture where there is an infinite set of Fock components.
The leading lightcone Fock states for highly boosted pions and nucleons are constructed from the ``plus'' component of the quark spinors~\cite{Burkardt:2002uc,Ji:2003yj}, while the overlap with standard interpolators includes a sub-leading lightcone wave function.

In this \emph{Letter}, we construct interpolators associated with the leading lightcone Fock states for pions and nucleons.
We study the Parisi-Lepage scaling~\cite{Parisi:1983ae,Lepage:1989hd} of correlation functions (correlators) built from these interpolators and show that their SNR is enhanced for large boosts $P$ by a kinematic factor proportional to $P^2$.
Proof-of-principle numerical calculations corroborate these theoretical expectations.
We find that although excited-state effects are larger for these new interpolators than standard ones for small boost values, excited-state effects are comparable or smaller for very large boosts.
To quantify SNR improvements in ground-state energy determinations, we use 
the Lanczos/Rayleigh-Ritz framework for correlator analysis~\cite{Wagman:2024rid,Hackett:2024xnx,Ostmeyer:2024qgu,Chakraborty:2024scw,Hackett:2024nbe,Abbott:2025yhm}, which
provides ground-state energy estimators with asymptotically constant SNR for boosted hadrons.
For the setup considered here, kinematically-enhanced interpolators lead to $\mathcal{O}(100)$-fold SNR improvement for pions with boosts of $P>2$~GeV and $\mathcal{O}(10)$-fold SNR improvement for nucleons with boosts of $P>3$~GeV.
Analogous SNR improvements are seen in three-point correlators, suggesting that these new interpolators can significantly improve the precision of lattice QCD studies of {boosted hadrons}. 

{\bf Kinematic enhancement: theory --- }
In the lightcone limit, a Dirac spinor can be decomposed into $\psi=\psi_++\psi_-$ with $\psi_\pm=\tfrac{1}{\sqrt{2}}\gamma_\mp\gamma_\pm\psi$, 
in which the ``plus'' component $\psi_+$ dominates the dynamics~\cite{Burkardt:2002uc,Ji:2003yj}.
Orienting the hadron momentum as $\vec{P} = |\vec{P}| \hat{e}_z$, the lightcone gamma matrices are $\gamma_\pm \equiv \tfrac{1}{\sqrt{2}}(\gamma_t\pm \mathbf{i}\gamma_z)$, where $\gamma_{x/y/z/t} \equiv \gamma_{1/2/3/4}$ are Euclidean gamma matrices satisfying $\gamma_\mu^\dagger = \gamma_\mu$ and $\{\gamma_\mu,\gamma_\nu\} = 2\delta_{\mu\nu}$ for $\mu,\nu=\{1,2,3,4\}$; Minkowski versions are related by $\gamma_4 =\gamma^M_0$ and $\gamma_i = \mathbf{i}\gamma^M_i$ for $i=\{1,2,3\}$.
Thus quark bilinears constructed with only $\psi_+$ components describe the leading Fock states of mesons with mass $M$ in the expansion of $M/(E+P_z)$. For pseudoscalar mesons like the pion, the leading contribution is $u_+^\dagger \gamma_5 d_+=\sqrt{2}\bar{u}\gamma_+ \gamma_5d$~\cite{Lepage:1979zb,Efremov:1979qk}. On the other hand, the traditional  pseudoscalar operator $\PiP =(u_+^\dagger\gamma_t \gamma_5 d_-+u_-^\dagger\gamma_t \gamma_5 d_+)/2$ is associated with subleading Fock states.
These lightcone physics considerations suggest that $\PiPmu$ operators should have better overlap with highly boosted pion ground states.

Axial-vector pion operators $\PiPmu$ transform differently from standard pseudoscalar pion operators $\PiP$ under rotations and only $\PiPt$ has non-zero overlap with the pion ground state in the rest frame.
Conversely, the axial-vector current $\PiAV$ has the same quantum numbers as an axial-vector meson and thus can be used to study the spectrum of mesons with axial-vector quantum numbers at $P_z=0$, including the $a_1(1260)$ resonance. 
However, rotational symmetry  no longer provides useful constraints for hadron states with non-zero $\vec{P}$, which allows $\PiAV$ to overlap with the same states as pseudoscalar pion operators.
This effect is familiar in the context of calculating the pion decay constant using the cross-correlation of $\PiPmu$ and $\PiP$, which has the spectral representation
\begin{equation}\label{eq:fpi}
    \begin{split}   
    &\sum_{\vec{x}}e^{\mathbf{i}\vec{P}\cdot \vec{x}} \langle[\bar{u}\gamma_\mu\gamma_5 d](\vec{x},t)[\bar{d}\gamma_5 u](0)\rangle\\
    &\qquad=\frac{e^{-E_\pi({\vec{P}})t}}{2E_\pi({\vec{P}})} Z_\pi({\vec{P}}) \mathbf{i}f_\pi P_\mu +\ldots,
    \end{split}
\end{equation}
where $f_\pi$ is the bare pion decay constant, defined from the ground-state pion-to-vacuum matrix element of the axial-vector current
$\langle \Omega | \bar{u}\gamma_\mu\gamma_5 d|\pi(\vec{P})\rangle=\mathbf{i}f_\pi P_\mu$, 
and $Z_\pi({\vec{P}}) \equiv \langle \pi({\vec{P}})|\bar{d}\gamma_5 u|\Omega\rangle$ is the ground-state overlap factor of the pseudoscalar pion operator. 
Once there is a large boost in the $z$-direction, the overlap of $\PiAV$ with moving pion {ground} states will be enhanced by $P_z$ as in Eq.~\eqref{eq:fpi} and dominate the signal.
{Some excited-state overlaps may also receive enhancements proportional to $P_z$; however, excited states can involve additional polarization and/or relative momentum vectors, which can lead to different relative enhancement of ground-state versus excited-state overlaps.}

The overlap between axial-vector and pseudoscalar operators follows from the symmetries of QCD in boosted frames.
In the continuum and infinite-volume limits, both pseudoscalar and axial-vector meson operators parallel to the momentum vector, $\PiAV$ for $\vec{P} = P_z \hat{e}_z$, transform trivially under the little group of rotations leaving $\vec{P}$ invariant.
The symmetries of the finite-volume lattice theory form a discrete subgroup of this little group, and $\PiP$ and $\PiAV$ transform in the same irreps of the lattice symmetry groups  $C_{4v}$ for boosted frames.
Explicitly, the pseudoscalar irrep  subduces as $A_1^- \rightarrow A_2$, and the axial-vector irrep subduces as $T_1^+ \rightarrow A_2 \oplus E$~\cite{Thomas:2011rh,Morningstar:2013bda,Detmold:2024ifm}.
It is precisely the $\PiAV$ component of {$(\bar{u}\gamma_i\gamma_5 d)$} that transforms in the $A_2$ irrep, confirming that $\PiAV$ and $\PiP$ transform identically.

Higher-spin interpolators overlap with lower-spin states in boosted frames only through extra $P_\mu$-dependent kinematic factors as in Eq.~\eqref{eq:fpi}.
At large momentum the overlap of $\PiPt$ with the pion ground state is related to that of the $\PiP$ operator by $f_\pi E_\pi / Z_\pi$,
{which provides an enhanced overlap of $\mathcal{O}(E_\pi)$  with boosted pion states.} A similar argument also applies to $\PiAV$, where the kinematic enhancement is {$\mathcal{O}(P_z)$}, except that this operator has zero overlap with the pion ground state in the rest frame.
In general, {the signal of $\PiPmu$ two-point correlators receives a kinematic enhancement of $\mathcal{O}(P_\mu^2)$} at large momentum.

{On the other hand, the statistical fluctuation of these correlation functions can be estimated through an analysis of variance correlators following the methods of Parisi and Lepage \cite{Parisi:1983ae,Lepage:1989hd}. }
For a generic pion interpolator $O_{\pi}(\vec{x},t)$, the variance correlator associated with $C_{\pi}(\vec{P},t) \equiv  \sum_{\vec{x}} O_{\pi}(\vec{x},t) O_{\pi}^\dagger(0) e^{\mathbf{i}\vec{P}\cdot\vec{x}}$,
\begin{equation}
\begin{split}
    \text{Var}(C_{\pi}) &= \left< \text{Re} (C_\pi)^2 \right> - \left< C_\pi \right>^2 \\
    &=  \frac{1}{2}\left< |C_\pi|^2 \right> + \frac{1}{2}\left<C_{\pi}^2\right> - \left< C_\pi \right>^2,
    \end{split}
\end{equation}
involves $C_{\pi}(\vec{P},t)^2$, which includes two-pion states with total momentum $2\vec{P}$, as well as
\begin{equation}
    |C_\pi|^2 
    = \sum_{\vec{x},\vec{y}}  e^{\mathbf{i}\vec{P}\cdot (\vec{x}-\vec{y})}O_\pi(\vec{y},t)O_\pi^\dagger(\vec{x},t)  O_\pi^\dagger(0)O_\pi(0),
\end{equation}
which has total momentum zero and is dominated at late time by states with two $|\vec{P}| = 0$ pions decaying at a rate $\sim 2m_\pi$, leading to exponential SNR decay for boosted pion states.

For these $|\vec{P}|=0$ two-pion states, the {$P_\mu^2$} factors arising for kinematically-enhanced interpolators are simply equal to {$m_\pi^2$} when they are non-zero.
At large $t$ the variance of $C_\pi$ is not kinematically enhanced, so $\text{SNR}(C_\pi) \equiv \left< C_\pi \right> / \sqrt{\text{Var}(C_\pi)}$ for $\PiPmu$ {receives an $\mathcal{O}(P_\mu^2/m_\pi^2)$ enhancement compared to $\PiP$ at large boost, as analyzed in Appendix \ref{app:quantitative}. Empirically, the precision of the traditional interpolator $\PiP$ is better than $\PiPt$ by an $\mathcal{O}(1)$ factor for static pions (in our test, this factor is around $2$), which can be easily compensated by the kinematic enhancement.} {Note that the enhancement should not depend significantly on the lattice spacing, because the improved interpolators have the same dimension as the traditional ones. Numerical results from applying these interpolators to a gauge-field configuration with $a \approx 0.06$ fm and $m_\pi \approx 670$ MeV are consistent with these expectations~\cite{Bollweg:2025iol}.}
Current state-of-the-art continuum-extrapolated parton structure calculations with near-physical pion masses have employed boosts of up to $P_z\sim 1.9$ GeV for parton distribution functions (PDFs)~\cite{Gao:2022iex} and up to $P_z \sim 2.2$ GeV for a determination of the Collins-Soper kernel~\cite{Avkhadiev:2024mgd}, corresponding to SNR enhancement factors for two-point correlators as large as $E_\pi^2/m_\pi^2 \sim 200$.

{Some of these interpolators have been explored in a generalized eigenvalue problem (GEVP) setup previously~\cite{Thomas:2011rh,Dudek:2012gj,Detmold:2022dmw} for static or slightly boosted meson systems. More recently, $\PiPt$ has been used along with $\PiP$ at larger boosts to calculate the pion distribution amplitude (DA)~\cite{Kovner:2024pwl},  but its SNR benefits were not revealed. One reason is that the enhancement factor for pion DA is linear in $E_\pi/m_\pi$, which was not large enough ($\lesssim5$) in those calculations to provide a noteworthy enhancement. Moreover, as we show below, $\PiPt$ has worse SNR than $\PiP$ for a static pion, which offsets its enhancement at small boosts.}

An identical strategy can be used to construct nucleon interpolators using the quark-field ``plus'' components.
The standard interpolator for a static nucleon is 
\begin{align}
    N_{\Gamma}=\epsilon_{abc}(d_a^TC\Gamma u_b)\mathcal{P}_+u_c,
    \label{eq:standard-nucleon-interpolator}
\end{align}
with $\Gamma=\gamma_5$ and $C$ the charge conjugate operator, $\mathcal{P}_\pm \equiv (1\pm\gamma_t)/2$ to project the positive-parity sector in the rest frame,
and $\epsilon_{abc}$ the Levi-Civita symbol with $a,b,c$ as color indices. 
Here, the diquark $(d_a^TC\gamma_5u_b)$ has spin 0. Conversely, the leading Fock component built from quark-field ``plus'' components has a spin-1 diquark~\cite{Braun:2000kw,Burkardt:2002uc,Ji:2003yj}.
Both $\Gamma=\gamma_5\gamma_\mu$ and $\Gamma=\gamma_\mu$ correspond to the same baryon leading-twist wave function in the lightcone limit
\cite{Zanotti:2003fx},
\begin{align}\label{eq:ncl_interpolator1}
    &\langle 0| N_{\gamma_5\gamma_\mu}|N(\vec{P})\rangle = \alpha P_\mu \mathcal{P}_+u(\vec{P})+\beta\gamma_\mu \mathcal{P}_+u(\vec{P}),\\
    &\langle 0| N_{\gamma_\mu}|N(\vec{P})\rangle =\alpha' P_\mu \gamma_5\mathcal{P}_+u(\vec{P})+\beta'\gamma_\mu \gamma_5\mathcal{P}_+u(\vec{P}), \nonumber
\end{align}
where $u(\vec{P})$ is the nucleon's Dirac spinor and $\alpha$, $\alpha'$, $\beta$, $\beta'$ are scalar functions of $\vec{P}$.
The term proportional to $P_\mu$ is not present in $\langle 0| N_{\gamma_5}|N(\vec{P})\rangle = Z(\vec{P}) \mathcal{P}_+u(\vec{P})$ with scalar $Z(\vec{P})$.
{Compared to the static nucleon, this term generates an enhancement $\propto P_\mu^2$ at large boost for the $N_{\gamma_5\gamma_\mu}$ and $N_{\gamma_\mu}$ two-point correlators that we estimate to be $\mathcal{O}(P_\mu^2/M_N^2)$ based on analogous arguments to those in Appendix~\ref{app:excited_state}. }
Both $N_{\gamma_5\gamma_\mu}$ and $N_{\gamma_\mu}$ overlap with spin-$\frac{3}{2}$ baryons, such as the $\Delta(1232)$~\cite{Zanotti:2002nk}, but the overlap is found below to be numerically small at large momentum.

In the rest frame, $N_{\Gamma}$ can also be projected by $\mathcal{P}_-$ to isolate the negative-parity sector containing the $N^*(1535)$ resonance. 
Boosted states are no longer eigen-states of the parity, thus any projection will contain states boosted from both parity sectors. Alternatives to the parity projector $\mathcal{P}_+$ in Eq.~(\ref{eq:standard-nucleon-interpolator}) are considered in Appendix~\ref{app:parity}, {where
we also found that the $\gamma_t$ term in the traditional projector $\mathcal{P}_\pm$ projects the quark with free spinor indices to its $u_+$ components and thus automatically enhanced the correlator at large boost compared to $\mathcal{P}=1$.}
In the remaining part of this work, we will show results only with $\mathcal{P}_+$ projection.

Parisi-Lepage analysis analogous to the pion case above shows that the boosted nucleon variance is dominated by states with three zero-momentum pions with energies $\sim 3 m_\pi$ that do not receive kinematic enhancements, and therefore the $N_{\gamma_5 \gamma_\mu}$ and $N_{\gamma_\mu}$ correlator SNR should receive $\mathcal{O}(P_\mu^2 / M_N^2)$ enhancement.

\begin{figure}
    \centering
    \includegraphics[width=0.99\linewidth]{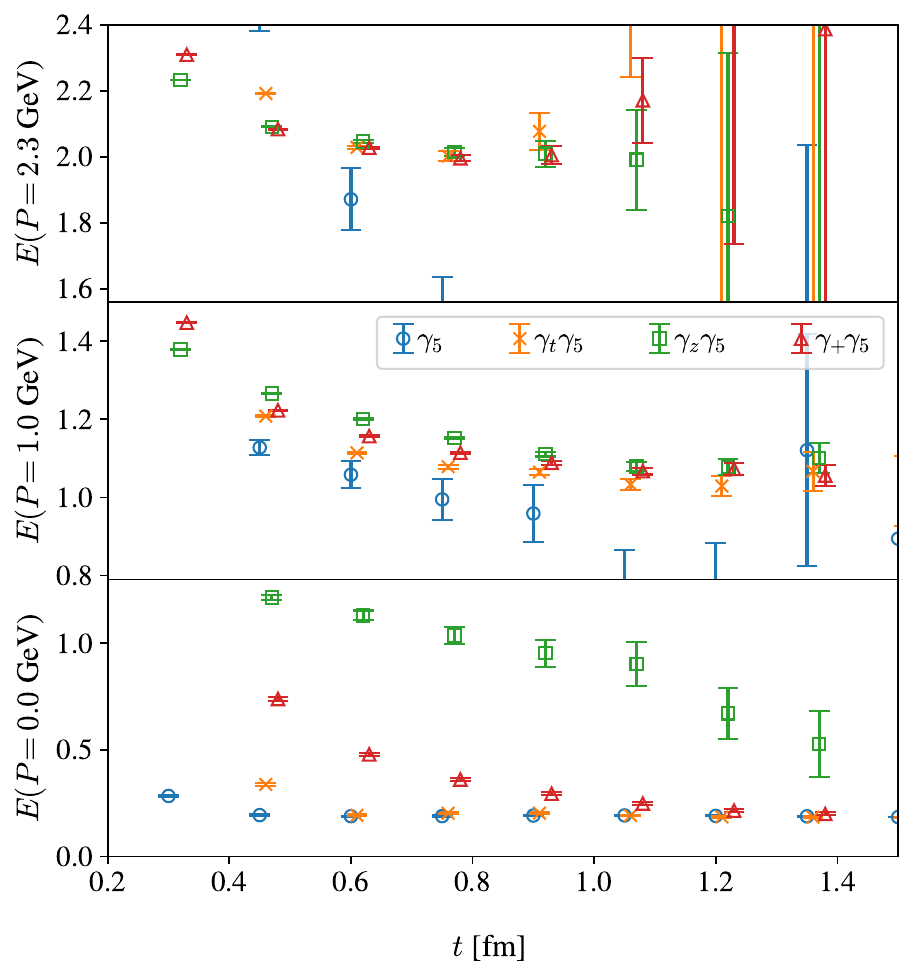}
    \caption{Comparison of the pion effective mass at various boosts with four interpolators. At large momentum, all three kinematically-enhanced interpolators $(\bar{u}\Gamma d)$ with $\Gamma=\{\gamma_t\gamma_5,\gamma_z\gamma_5,\gamma_+\gamma_5\}$ show substantial precision improvements compared with the traditional $(\bar{u}\gamma_5 d)$ interpolators.}
    \label{fig:pion_eff_mass}
\end{figure}

{\bf Lattice QCD verification --- }
We test these new interpolators numerically on a gauge ensemble produced by the MILC collaboration~\cite{MILC:2012znn} with 2+1+1 flavors of highly improved staggered quarks (HISQ) tuned to reproduce the physical pion mass and the one-loop Symanzik improved gauge action~\cite{Symanzik:1983dc}.   The lattice has a volume $L^3\times T=32^3\times48$ and lattice spacing $a\approx0.15$~fm. We apply two steps of HYP smearing with parameters $\{\alpha_1,\alpha_2,\alpha_3\}=\{0.75,0.6,0.3\}$~\cite{Hasenfratz:2001hp} to the gauge fields and then use a Wilson-clover action~\cite{Sheikholeslami:1985ij} for the valence quarks with 
$c_\text{SW} = 1$ and $\kappa=0.12635$,
tuned to produce $\approx 190$~MeV pions.  We perform measurements with 64 source locations on 334 configurations. To increase the signal at large momentum, we use momentum smearing of $k\approx1.55$~GeV for pion momentum from $P_z=0$ to $P_z=2.32$~GeV~\cite{Bali:2016lva}. 

Figure~\ref{fig:pion_eff_mass} confirms significant SNR improvement in the effective masses $E_{\rm eff}(t,P) \equiv 1/a  \ln \left[ C(t,P)/C(t+a,P) \right]$ of $\PiPmu$ interpolators at large momenta. It is clear that $\PiAV$ only overlaps with heavier states at $P_z=0$ but becomes dominated by pion states at large momenta. At the largest momentum, $P_z\sim 2.32$~GeV, $\PiP$ correlators become very noisy for $t \gtrsim 0.6$~fm and decrease to unphysical values, which is a clear sign of noise dominance; however, $\PiPmu$ interpolators can achieve reliable signals out to larger $t \sim 0.9$~fm. 

Besides $\PiAV$ and $\PiPt$, we compute correlators for $\PiPlus$ that can be directly associated with the leading lightcone Fock states of the pion. When the two interpolators $(\bar{u}\gamma_t\gamma_5 d)$ and $(\bar{u}\gamma_z\gamma_5 d)$ have equal quality, there is an extra factor-of-$2$ enhancement in the signal for $(\bar{u}\gamma_+\gamma_5 d)$, while the noise does not increase as much due to their correlations, resulting in a slightly better SNR than either component.
In general, we observe the SNRs follow the ordering ${\rm SNR}(\bar{u}\gamma_+\gamma_5 d)>{\rm SNR}(\bar{u}\gamma_z\gamma_5 d)>{\rm SNR}(\bar{u}\gamma_t\gamma_5 d)$ at large momenta. 

\begin{figure}[t]
    \centering
    \includegraphics[width=0.99\linewidth]{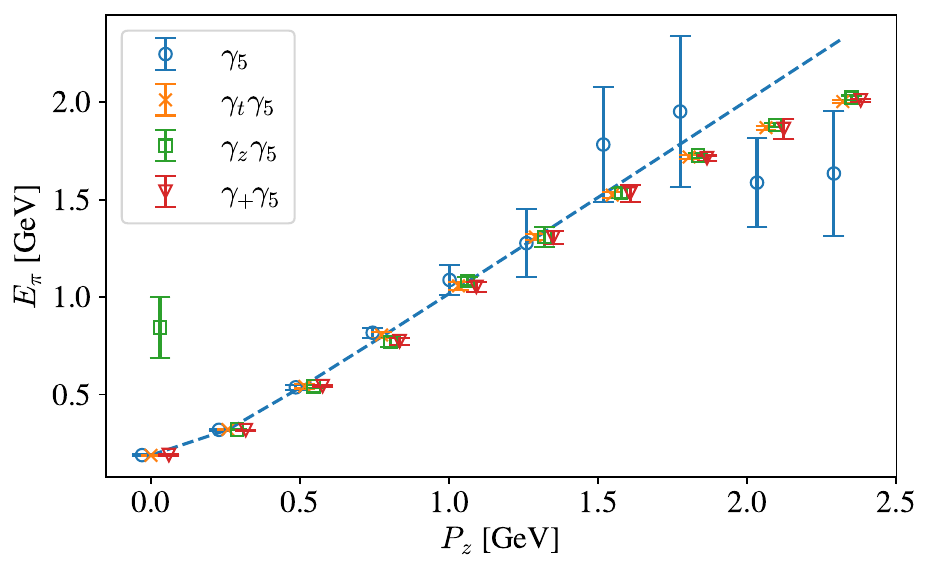}
    \caption{The ground-state energy of the pion extracted using the Lanczos/Rayleigh-Ritz method at various boosts for the interpolators $(\bar{u}\Gamma d)$ with $\Gamma=\{\gamma_5,\gamma_t\gamma_5,\gamma_z\gamma_5,\gamma_+\gamma_5\}$. {The dashed line represents the continuum dispersion relation.} The precision is substantially improved at large momentum when using the kinematically-enhanced $(\bar{u}\gamma_\mu\gamma_5 d)$. Note that for systems at rest, $(\bar{u}\gamma_z\gamma_5 d)$ has zero overlap with the pion.  }
    \label{fig:pion_fit}
\end{figure}

{As shown in Fig.~\ref{fig:pion_eff_mass},} at small momentum, $\PiPmu$ correlators converge slower than $\PiP$, indicating larger excited-state effects. At large momentum,
$\PiPmu$ converges faster. In a na\"ive implementation, the operators $\PiAV$ and $\PiPlus$ work best for large momentum, but $\PiPt$ is more suitable to scan a large range of momenta. 
A strategy for removing excited-state contamination from $\PiAV$ correlation functions by forming differences with correlation functions built from a transverse component, e.g., $(\bar{u}\gamma_x\gamma_5 d)$, is presented in Appendix~\ref{app:excited_state}.

We extract the ground-state energy $E_\pi(P_z)$ from the two-point correlators with the Lanczos/Rayleigh-Ritz method~\cite{Wagman:2024rid,Hackett:2024xnx,Ostmeyer:2024qgu,Chakraborty:2024scw,Hackett:2024nbe,Abbott:2025yhm} using nested bootstrap median estimators and spurious-state filtering with the ZCW test with $F_{\rm ZCW} = 10$~\cite{Hackett:2024nbe}. 
To avoid the effects from extra terms arising from Wick's theorem at $t=0$, as explained in Appendix~\ref{sec:t=0}, we start with $t=2a$ data.
Lanczos then converges within a few iterations; 
results after 22 iterations, incorporating $t \in [2, N_t - 3]a$, are shown in Fig.~\ref{fig:pion_fit}. {The results deviate from the continuous dispersion relation up to $15\%$ at very large momentum due to discretization effects, as has been shown in Ref.~\cite{Avkhadiev:2024mgd}.}
For $P_z=0$, the ground state identified from $\PiAV$ correlators is significantly heavier than the pion mass, as expected due to its $T_1^+$ quantum numbers.
For non-zero momenta, all correlators provide statistically consistent ground-state energy estimates.
As the momentum increases, the kinematically-enhanced interpolators clearly show growing SNR improvement compared to the traditional $\PiP$ interpolator.
The improvement is consistent with $\mathcal{O}(P_\mu^2 / m_\pi^2)$ scaling for all but the smallest momenta and reaches factors of $\sim 30$-$50$ for $P_z>2$~GeV. Appendix~\ref{app:quantitative} provides another way to estimate the enhancement as a function of $P_z/m_\pi$, where we have also included data with a heavier pion mass $m_\pi\approx 400$~MeV to show the scaling. It demonstrates an improvement factor of up to $\sim 50$ that is consistent with the Lanczos analysis.

\begin{figure}[t]
    \centering
    \includegraphics[width=0.99\linewidth]{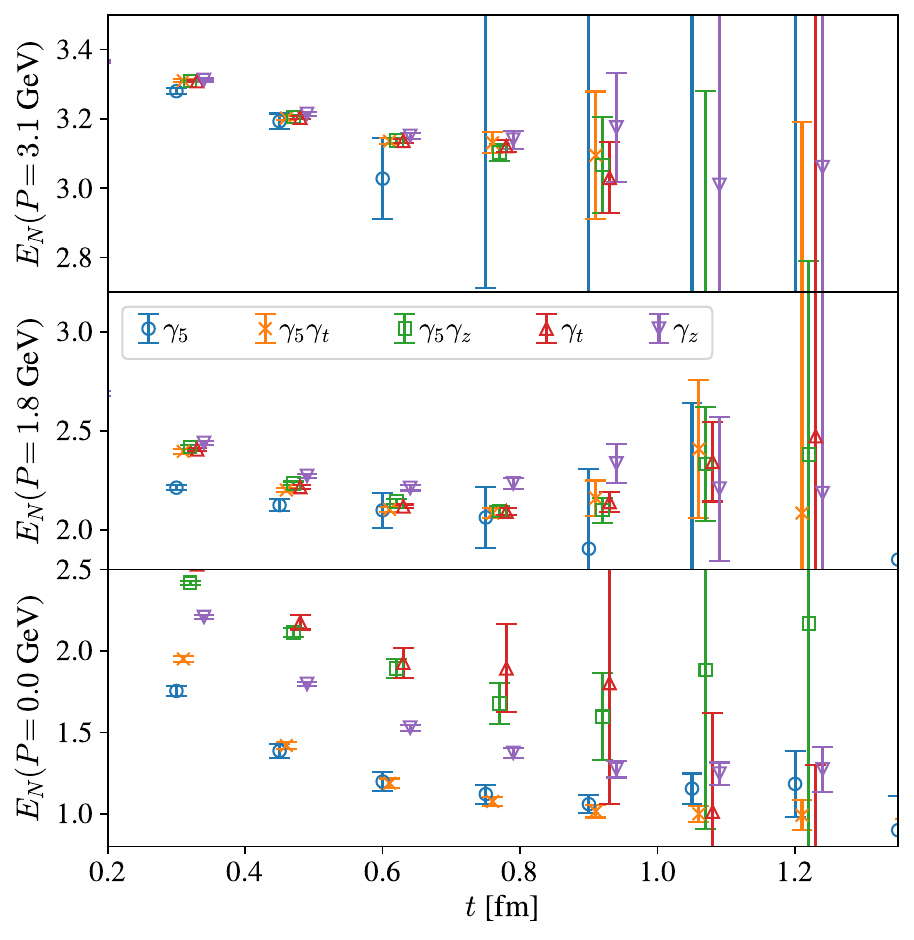}
    \caption{ The analog of Fig.~\ref{fig:pion_eff_mass} for the nucleon, where the interpolating operators are $(d^T C\Gamma u)\mathcal{P}_+u$ with $\Gamma=\{\gamma_5,\gamma_5\gamma_t,\gamma_5\gamma_z,\gamma_t,\gamma_z\}$. At large momentum, the last four operators all yield higher precision than the choice of $\Gamma=\gamma_5$.}
    \label{fig:ncl_eff_mass_ops}
\end{figure}

We measure the nucleon two-point correlators on the same lattice with 16 sources and 202 configurations. To reach higher momentum, 
we choose $P=\frac{2\pi n}{L}\times(1,1,1)$ for $n\in[0,7]$, i.e.~up to $3.1$~GeV. 
The momentum smearing is optimized for the largest momentum $P=3.1$~GeV. 
Figure~\ref{fig:ncl_eff_mass_ops} shows the effective mass for static and boosted correlators with the five different 
$N_\Gamma$'s
where $\Gamma \in \{\gamma_5, \gamma_5 \gamma_t, \gamma_5 \gamma_z, \gamma_t, \gamma_z \}$.\footnote{
For a general boost with momentum $\vec{P}$ not aligned with the $z$-axis, the spatial gamma matrix becomes $\vec{\gamma} \cdot \hat{P}$. In our case, where $\vec{P} \propto (1,1,1)$, the projector is $\Gamma = (\gamma_x + \gamma_y + \gamma_z)/\sqrt{3}$, which we denote as the $z$-component for simplicity.
}
All kinematically-enhanced interpolators show similar SNR improvements compared to $N_{\gamma_5}$ at large $P$. Among them, $N_{\gamma_5\gamma_t}$ has the least excited-state contamination where differences are visible.

Unlike the pion case where the pseudoscalar interpolator is optimal in the rest frame, the SNR for the nucleon interpolator $N_{\gamma_5\gamma_t}$ is equally good as $N_{\gamma_5}$ at $P=0$.
At large momentum, the enhancement factor is of similar size as $E_N^2/M_N^2$ without the $\mathcal{O}(1)$ loss that appears in the pion case. 
Lanczos analysis of the nucleon correlators with the same spurious-state filtering as above gives a ground-state energy estimator with SNR enhanced by factors of 3--10 for the kinematically-enhanced interpolators in comparison with $N_{\gamma_5}$.
With larger $\epsilon_{\rm ZCW}$, Lanczos analysis of $N_{\gamma_5}$ leads to more precise signals but of a higher-energy state than that resolved from the kinematically-enhanced interpolators.

\begin{figure}[t]
    \centering
    \includegraphics[width=0.99\linewidth]{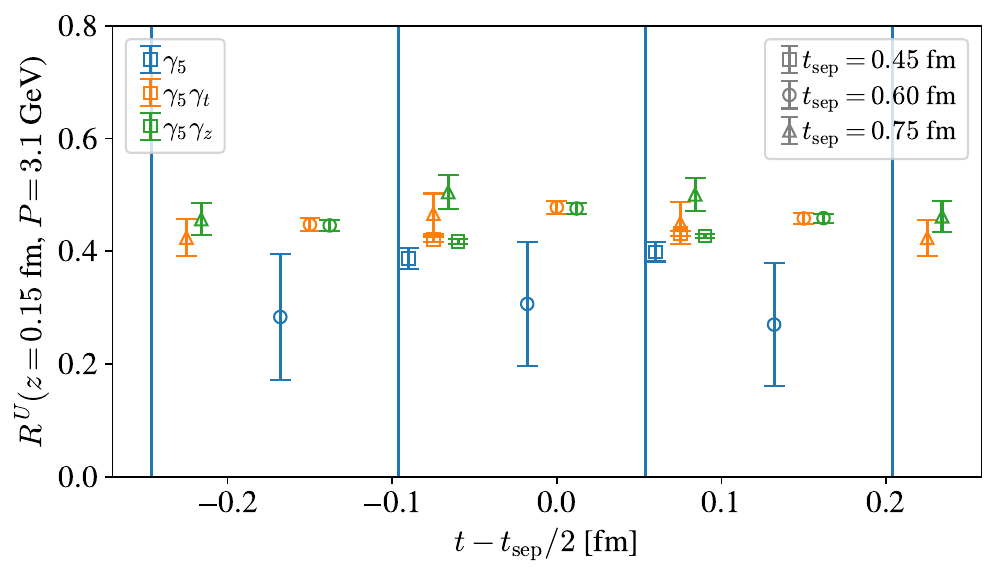}
    \caption{A ratio of three-point to two-point functions using three nucleon interpolators from Fig.~\ref{fig:ncl_eff_mass_ops} shows improved precision with kinematically enhanced choices $\Gamma = \{\gamma_5\gamma_t, \gamma_5\gamma_z\}$ over the standard $\Gamma = \gamma_5$.}
    \label{fig:ncl_3pt_ratio}
\end{figure}

The kinematic enhancement applies exactly the same way in the three-point correlators. For an illustration, we measure the bare unpolarized quark quasi-PDF matrix element~\cite{Ji:2013dva,Gao:2023lny} {$h_B^U(z,P_z)$ of the nucleon for up quark connected diagrams in the Coulomb gauge,
\begin{align}
    C_{\rm 3pt}&(\vec{P},t,t_{\rm sep}) =\sum_{\vec{x},\vec{w},\vec{y}} e^{\mathbf{i} (\vec{x}-\vec{y})\cdot \vec{P}} \text{Tr}\left[ \vphantom{\frac{1}{2}} \mathcal{P}_+ \langle  N_\Gamma(t_{\rm sep},\vec{x}) \right. \nonumber\\
    &\hspace{20pt} \left. \times \overline{u}(t,\vec{w}+z\hat{z})\gamma_t u(t,\vec{w})\  \overline{N}_\Gamma(0,\vec{y})\rangle \vphantom{\frac{1}{2}} \right],
\end{align}
}
which has the same kinematic enhancement as the two-point correlators. Thus we take the ratio 
\begin{align}
    R^U(t,t_{\rm sep})=\frac{C_{\rm 3pt}(t,t_{\rm sep})}{C_{\rm 2pt}(t_{\rm sep})} ={h^U_B(z,P_z)}+\dots,
\end{align}
where the omitted terms are contamination from excited states. {We show the results at $z=a$ and $P=3.1$~GeV in Fig.~\ref{fig:ncl_3pt_ratio}}.
The results are consistent among the three interpolators, and we clearly observe a significant improvement with the new interpolators at the level of correlators, especially for larger $t_{\rm sep}$ where the ground states are dominating.

{\bf Conclusion --- }
In this work, we propose interpolating operators for lattice QCD calculations of highly-boosted pions and nucleons
with kinematically-enhanced ground-state overlap factors at large momentum. The signal of lattice correlators is enhanced quadratically in the Lorentz boost factor, while the noise is insensitive to the momentum, resulting in a kinematically-enhanced SNR. Compared to the traditional interpolators, we find an improvement in the SNR by up to $\sim50$ for pions with $P_\mu^2/m_\pi^2\approx150$, and up to $\sim10$ for nucleons with $P_\mu^2/M_N^2\approx10$, which correspond to increases of statistics by $ \mathcal{O}(2000)$ and $ \mathcal{O}(100)$, respectively. 
Using these interpolators will tremendously reduce the cost of measuring boosted hadron spectra and matrix elements, significantly improving the precision of lattice calculations of form factors at large $Q^2$ and partonic observables. Such high-precision, high-momentum lattice calculations of form factors and parton distributions 
are necessary inputs for analyzing collider experiments, including the LHC and the upcoming Electron-Ion Collider. Moreover, they could potentially be extended to processes such as $\pi\pi$ scattering~\cite{Blanton:2021llb,RBC:2021acc} and heavy meson decays to energetic final states, which are essential to resolve CP matrix elements for high-precision unitary violation searches.

\begin{acknowledgments}
We thank Artur Avkhadiev, Yang Fu, Jinchen He, Xiangdong Ji, Luchang Jin, Andreas Kronfeld, Andreas Sch\"{a}fer, Yushan Su, and Ruth Van de Water for valuable discussions.
This material is based upon work supported by the U.S. Department of Energy, Office of Science, Office of Nuclear Physics through Contract No.~DE-AC02-06CH11357, the Scientific Discovery through Advanced Computing (SciDAC) award \textit{Fundamental Nuclear Physics at the Exascale and Beyond}, the Quark-Gluon Tomography (QGT) Topical Collaboration under contract no.~DE-SC0023646, and the Fermi Research Alliance, LLC under Contract No.~DE-AC02-07CH11359 with the U.S. Department of Energy, Office of Science, Office of High Energy Physics.
Argonne National Laboratory's contribution is also based upon work supported by Laboratory Directed Research and Development (LDRD) funding from Argonne National Laboratory, provided by the Director, Office of Science, of the U.S. Department of Energy under Contract No.~DE-AC02-06CH11357.
We gratefully acknowledge the computing resources provided on Swing, a high-performance computing cluster operated by the Laboratory Computing Resource Center at Argonne National Laboratory. This research used resources of the Argonne Leadership Computing Facility, a U.S. Department of Energy (DOE) Office of Science user facility at Argonne National Laboratory and is based on research supported by the U.S. DOE Office of Science-Advanced Scientific Computing Research Program, under Contract No. DE-AC02-06CH11357. This research also used facilities of the USQCD Collaboration, which are funded by the Office of Science of the U.S. Department of Energy. Our calculation is performed using the GLU~\cite{Hudspith:2014oja} and QUDA~\cite{Clark:2009wm} software packages.

\end{acknowledgments}

\appendix
\section{Quantitative analysis of the kinematic enhancement}\label{app:quantitative}
\begin{figure}[t]
	\centering
	\includegraphics[width=0.99\linewidth]{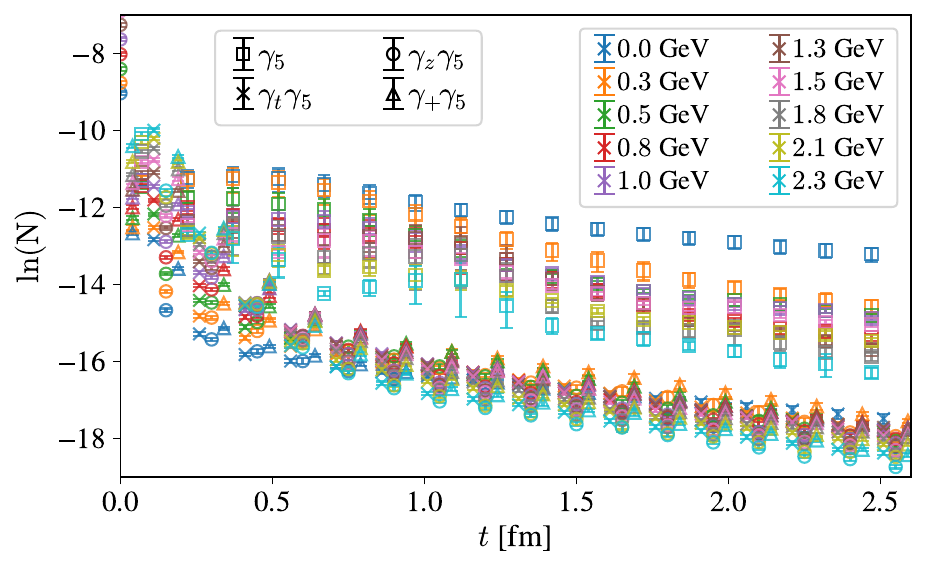}
	\caption{The scaling of the noise ${\rm N}$, defined as the standard deviation of the correlation function, as a function of source-sink separation at various momenta. In all cases, ${\rm N}\propto e^{-m_\pi t}$ asymptotically, but the prefactor is much smaller for the improved interpolators $ (\bar{u}\gamma_\mu\gamma_5 d)$.}
	\label{fig:pion_noise}
\end{figure}

\begin{figure}[t]
	\centering
	\includegraphics[width=0.99\linewidth]{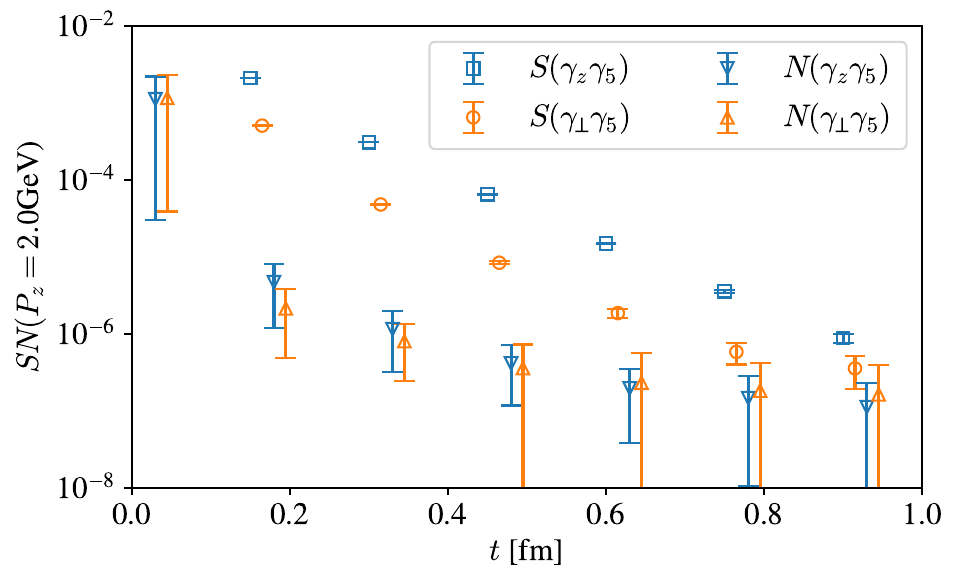}
	\caption{A comparison of the signal $\rm{S}$ and noise $\rm{N}$, defined as the mean value and standard deviation of the correlation function, for the meson boosted in the $z$-direction using interpolators $(\bar{u}\gamma_z\gamma_5 d)$ and $(\bar{u}\gamma_x\gamma_5 d)$. Both have comparable noise, but the $z$-aligned interpolator leads to a substantial enhancement in the signal.}
	\label{fig:axial_noise}
\end{figure}

To quantify the improvement of the new pion interpolators, in principle we need to take a ratio of the SNR among them.
However, since the signal of the traditional interpolator $(\bar{u}\gamma_5 d)$ quickly decays to the noise-dominant region, its SNR becomes just a constant $\mathcal{O}(1)$ fluctuation. Thus it is not a faithful comparison beyond $0.45$~fm for large momentum. At this early Euclidean time, the correlators may not yet be dominated by the ground state pion, so a direct SNR comparison will not accurately reflect the enhancement in pion. However, we notice that the noise ${\rm N}(\bar{u}\Gamma d)$ follows a nice asymptotic $\cosh$-like behavior with the ground-state pion mass, as shown in Fig.~\ref{fig:pion_noise}. The kinematic enhancement only exists in the signal but not in the noise, as shown by a comparison between $(\bar{u}\gamma_z\gamma_5 d)$ and $(\bar{u}\gamma_x\gamma_5 d)$ in Fig.~\ref{fig:axial_noise}. In both plots, we estimate the uncertainty of the noise from the variance of the noise calculated in each jackknife sample with $n-1$ configurations. Since the fluctuations in the noise are more stable at large $t$, a quantitative comparison is more reliable for the noise measurements themselves. If we can match the corresponding signals to the same level, then the comparison of noise will be equivalent to the comparison of the actual SNR.

To realize this goal, we utilize the fact that the ground state signal falls off in the same asymptotic form among different interpolating operators, except for $\PiAV$ at rest. Thus, on a logarithmic scale, they just differ by a vertical shift at large $t$. 
More specifically, the shift can be estimated through the partially conserved axial current (PCAC) relation~\cite{Gell-Mann:1960mvl},
\begin{align}
	\langle \pi|\bar{d}\gamma_5u|\Omega\rangle\approx\frac{1}{m_l}\partial^\mu\langle \pi|\bar{d}\gamma_\mu\gamma_5u|\Omega\rangle\approx\frac{\mathbf{i}f_\pi m^2_\pi}{m_l},
\end{align}
where $m_l$ is the light quark mass, indicating that asymptotically,
\begin{align}
	{\rm S}(\bar{u}\gamma_5 d)\approx\frac{m_\pi^4}{m_l^2P_\mu^2}{\rm S}(\bar{u}\gamma_\mu\gamma_5 d).
\end{align}

{
Although the relation is inexact, we can use a one-parameter model to approximate the data,
\begin{align}
	{\rm S}(\bar{u}\gamma_5 d)\approx {\rm S}'(\bar{u}\gamma_\mu\gamma_5 d)\equiv {\rm S}(\bar{u}\gamma_\mu\gamma_5 d) \times \frac{\lambda}{P_\mu^2},
\end{align}
where the free parameter $\lambda\approx\frac{m_\pi^4}{m_l^2}$ is a mass-dimension two parameter that can in principle be momentum-, smearing- and quark-mass-dependent.
Near the chiral limit, $\lambda$ is an $\mathcal{O}(\Lambda_{\rm QCD}^2)$ constant according to the Gell-Mann-Oakes-Renner relation (often denoted $4B^2$), up to higher-order corrections computable in chiral perturbation theory.
We find that with a fixed $\lambda\approx 1.5/a^2$, the scaled correlators ${\rm S}'(\bar{u}\gamma_\mu\gamma_5 d)$ are consistent with ${\rm S}(\bar{u}\gamma_5 d)$ before noise dominance at all momenta, as shown in Fig.~\ref{fig:pion_signal_rescale}.
\begin{figure}[t]
	\centering
	\includegraphics[width=0.99\linewidth]{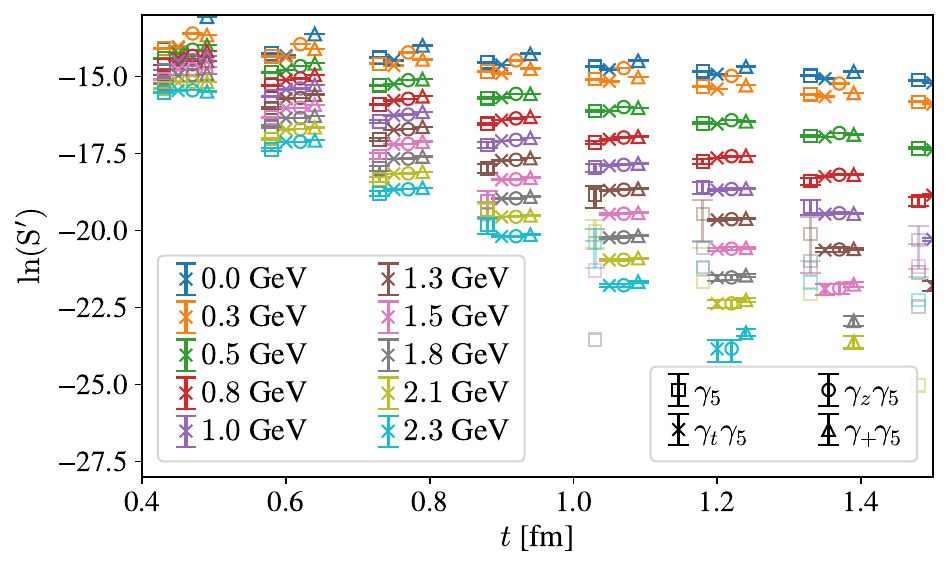}
	\caption{The rescaled signal ${\rm S}'$  for four pion correlators with different interpolators, defined as the signal ${\rm S}$ multiplied by a time-independent factor such that ${\rm S}'$ is comparable for all interpolators at early time. At fixed momentum, all correlators decay at the same rate, so ${\rm S}'$ remains independent of the choice of interpolator even at late time, until the point where the signal is lost to noise.}
	\label{fig:pion_signal_rescale}
\end{figure}
Then we can quantitatively compare the SNR by taking a ratio of their noises including the factor $\lambda /P_\mu^2$,
\begin{equation}
	\frac{{\rm SNR}(\bar{u}\gamma_\mu\gamma_5 d)}{{\rm SNR}{(\bar{u}\gamma_5 d)}}\approx \frac{{\rm N}'{(\bar{u}\gamma_5 d)}}{{\rm N}'{(\bar{u}\gamma_\mu\gamma_5 d)}}\equiv \frac{P_\mu^2}{\lambda}\frac{{\rm N}{(\bar{u}\gamma_5 d)}}{{\rm N}{{(\bar{u}\gamma_\mu\gamma_5 d)}}}.
\end{equation}
Note that the variance correlator ground-state is two pions at rest, thus the sizes of the noise would be proportional to the overlaps of these operators with a static pion. The numerator ${\rm N}{(\bar{u}\gamma_5 d)}$ is dominated by the overlap of the spinor structure $\bar{u}\gamma_5 d=(u_-\gamma_t\gamma_5 d_++u_+\gamma_t\gamma_5 d_-)/2$ with the static pion, while the denominator ${\rm N}{{(\bar{u}\gamma_\mu\gamma_5 d)}}$ comes from only the fluctuation of the plus components $u^\dagger_+\gamma_5 d_+$. This indicates that the latter has a more suppressed fluctuation, ${\rm N}{(\bar{u}\gamma_5 d)}/{\rm N}{{(\bar{u}\gamma_\mu\gamma_5 d)}}\propto {\lambda}/{m_\pi^2}$ up to an $\mathcal{O}(1)$ factor. As a result, 
\begin{align}
    \frac{{\rm SNR}(\bar{u}\gamma_\mu\gamma_5 d)}{{\rm SNR}{(\bar{u}\gamma_5 d)}}\propto \frac{P_\mu^2}{m_\pi^2}.
\end{align}
}

\begin{figure}[t]
	\centering
	\includegraphics[width=0.99\linewidth]{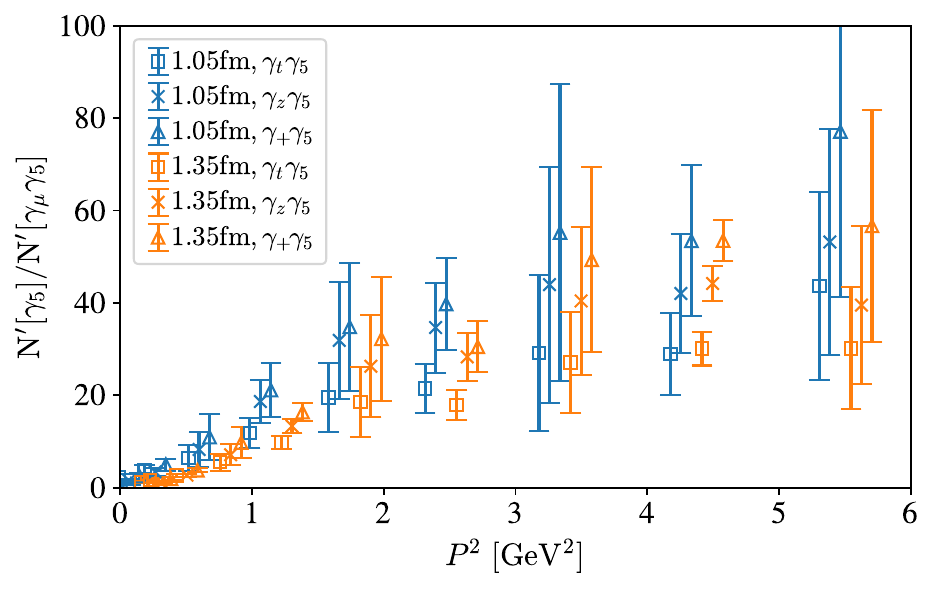}
	\caption{The ratio of rescaled noises, ${\rm N}'(\bar{u}\gamma_5 d)/{\rm N}'(\bar{u}\gamma_\mu\gamma_5 d)$, for $190$~MeV pion interpolators. We can achieve noise reduction by a factor of $\sim50$ at $P_z\approx 2.32$~GeV, which corresponds to an $\mathcal{O}(2000)$-fold increase in statistics. The $(\bar{u}\gamma_+\gamma_5 d)$ performs best among the axial vector interpolators at large momenta. }
	\label{fig:pion_noise_reduction}
\end{figure}

Figure~\ref{fig:pion_noise_reduction} shows the ratio of rescaled noises for two large time slices, 
which is a good approximation to ${\rm SNR}(\bar{u}\gamma_\mu\gamma_5 d)/{\rm SNR}(\bar{u}\gamma_5 d)$. The improvement clearly grows in an almost-linear pattern with $P^2$. At the largest momentum $P_z=2.3$~GeV, the improvement in signal-to-noise ratio can be a factor as large as 40 to 50, consistent with the Lanczos results, corresponding to an increase of statistics by a factor of $\mathcal{O}(2000)$. The final enhancement factor is roughly $1/3$ of the kinematic factor $P_z^2/m_\pi^2$ due to the loss of precision with the $(\bar{u}\gamma_t\gamma_5 d)$ operator compared to the $(\bar{u}\gamma_5 d)$ in the rest frame. Note that this is still for a heavier pion of mass around $190$~MeV, which is already $1.4$ times the physical value. Taking this factor into account, the improvement on phyiscal pion measurements can potentially reach a factor of $\mathcal{O}(10^4)$ increase in statistics at the same momentum.

\begin{figure}[t]
	\centering
	\includegraphics[width=0.99\linewidth]{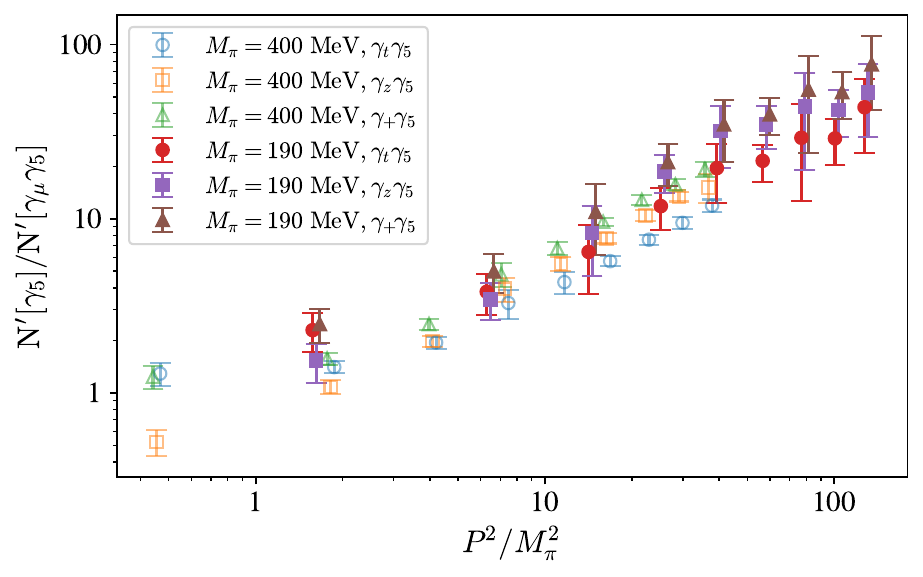}
	\caption{The ratio of noise between the rescaled noise,  ${\rm N}'(\bar{u}\gamma_5 d)/{\rm N}'(\bar{u}\gamma_\mu\gamma_5 d)$, compared to the predicted enhancement factor $P^2/M^2_\pi$. This relation is linear on a log-log plot, indicating that the predicted enhancement is correct up to a constant $\mathcal{O}(1)$ factor. Notably, the data at two different pion masses ($190$~MeV and $400$~MeV) lie on the same line, giving support to the theoretical prediction that the enhancement factor scales inversely with $m_\pi^2$.}
	\label{fig:pion_noise_reduction_heavy}
\end{figure}

To confirm that the kinematic enhancement works better for lighter pion masses and justify the empirical claim that $\lambda$ is approximately quark-mass independent  away from the chiral limit, we perform the same measurement with valence pion mass $m_\pi\approx 400$~MeV, roughly twice that of the previous test. Following the same procedure, we find the scaling in $P_z^2/m_\pi^2$ to be the same as the lighter pion case, and the enhancement factor to be about 4 times smaller at the same $P_z$, as shown in Fig.~\ref{fig:pion_noise_reduction_heavy}. 

\section{Baryon interpolator spin projections}\label{app:parity}

According to the analysis above, the $\psi_+$ component of the spinor $\psi$ is kinematically enhanced in a boosted frame. The same enhancement also applies to the free quark spinor in the nucleon interpolator. {In the spin contraction of the nucleon two-point correlator, the projection operator always appears in the following structure (we have suppressed all the color structures):
\begin{align}
	\langle N|N\rangle=&c_1{\rm Tr_s}[\mathcal{P}S_U]{\rm Tr_s}[S_U\Gamma S^T_D \Gamma']\nonumber\\
    &+c_2{\rm Tr_s}[S_U\mathcal{P}S_U\Gamma S^T_D\Gamma'],
\end{align}
where ${\rm Tr_s}$ labels the trace over spin indices, $S_{U/D}$ are the up/down quark propagators labeling the three quark fields in the nucleon interpolator, and the Dirac matrices $\Gamma$ and $\Gamma'$ are determined by the diquark structure. The quark propagator can be re-written as the Wick contraction of the spinor fields $S_U\sim u\bar{u}$. This indicates that there is always a sub-structure $\bar{u}\mathcal{P}u$ in both ${\rm Tr_s}[\mathcal{P}S_U]$ and ${\rm Tr_s}[S_U\mathcal{P}S_U\Gamma S^T_D\Gamma']$,
which has the same spinor structure as a meson interpolator $\bar{u}\Gamma u$. Thus $\mathcal{P}=\gamma_t$ will project out the $u_+$ component of the two adjacent quark fields $\bar{u}\gamma_tu=(u^\dagger_+u_++u^\dagger_-u_-)/\sqrt{2}$, introducing a kinematic enhancement factor of $E/M$ to the correlator. As a result, the $\mathcal{P}=\gamma_t$ projection at large momentum becomes more precise than the $\mathcal{P}=1$ projection. }
\begin{figure}[t]
	\centering
	\includegraphics[width=0.99\linewidth]{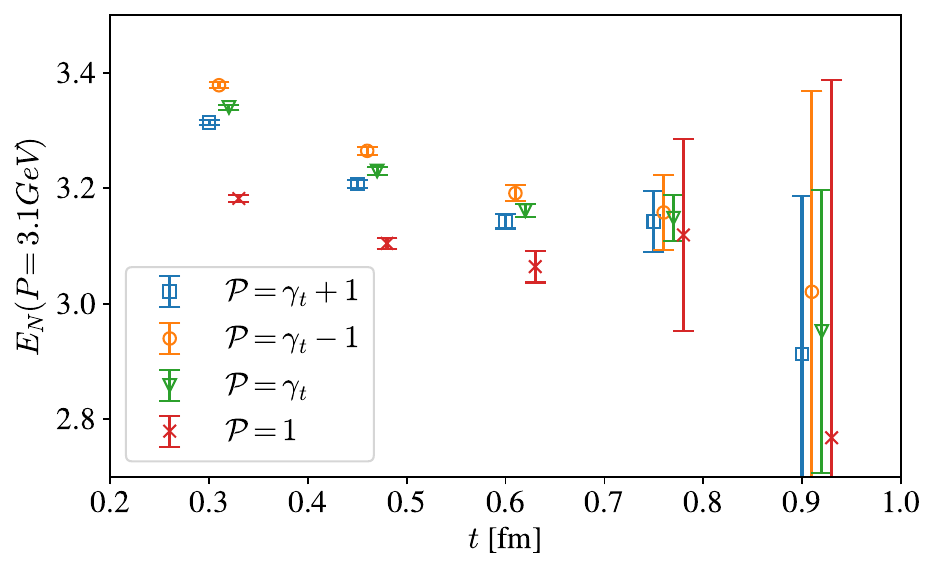}
	\caption{The effective mass plot of the nucleon two-point correlators at a boost of $3.1$ GeV using various parity projectors.}
	\label{fig:parity}
\end{figure}

Figure~\ref{fig:parity} shows the comparison among different projecting operators. 
One can tell that the projection operators with a $\gamma_t$ component are cleaner than $\mathcal{P}=1$. Since $\gamma_t=\mathcal{P}_+-\mathcal{P}_-$ comes from a linear combination of two different parity projections in the rest frame, it contains more contamination from negative parity states than $\mathcal{P}_+$ but less than $\mathcal{P}_-$. 

Note that the $\mathcal{P}_+$ projection is positive definite by definition of the lattice correlator, while  $\mathcal{P}_-$ is negative definite. So only $\mathcal{P}_+$, $-\mathcal{P}_-$, and $\gamma_t=\mathcal{P}_+-\mathcal{P}_-$ project out positive-definite correlators, where all excited states contribute with the same sign. On the other hand, $1=\mathcal{P}_++\mathcal{P}_-$ projection contains excited state contribution with indefinite signs thus may result in a fake plateau in the effective mass, the same as in correlators with asymmetric source-sink interpolators. 
One needs to be more careful when analyzing correlators with this projection.

\section{Excited-state suppression for the pion}\label{app:excited_state}

In the continuum, infinite-volume limit, there are two sets of excited-state contamination in the higher-spin interpolators. For example, the two-point correlator of two axial-vector interpolators contains two different spin components,
\begin{align}
	\langle [\bar{u}\gamma_\mu\gamma_5 d]^\dagger [\bar{u}\gamma_\nu\gamma_5 d]\rangle=&\sum_{n,S=1} \left(\frac{P_\mu P_\nu}{M_{n,1}^2}-g_{\mu\nu}\right)C_{n,1} \nonumber\\
	&+\sum_{n,S=0} \frac{P_\mu P_\nu}{M_{n,0}^2}C_{n,0},
\end{align}
where $M_n$ is the invariant mass of the $n$th state. Taking both indices to be $\{t,z,\perp\}$, we find that
\begin{align}
	\langle [\bar{u}\gamma_t\gamma_5 d]^\dagger [\bar{u}\gamma_t\gamma_5 d]\rangle&=\sum_{n,S=1} \frac{P_z^2}{M_{n,1}^2}C_{n,1} 
	+\sum_{n,S=0} \frac{E^2}{M_{n,0}^2}C_{n,0}, \nonumber\\
	\langle [\bar{u}\gamma_z\gamma_5 d]^\dagger [\bar{u}\gamma_z\gamma_5 d]\rangle&=\sum_{n,S=1} \frac{E^2}{M_{n,1}^2}C_{n,1} 
	+\sum_{n,S=0} \frac{P_z^2}{M_{n,0}^2}C_{n,0},\nonumber\\
	\langle [\bar{u}\gamma_\perp\gamma_5 d]^\dagger [\bar{u}\gamma_\perp\gamma_5 d]\rangle&=\sum_{n,S=1} C_{n,1}.
\end{align}
Compared to the ground-state enhancement, at very large momentum, both $(\bar{u}\gamma_t\gamma_5 d)$ and $(\bar{u}\gamma_z\gamma_5 d)$ correlators have suppressed excited-state contamination of order $\frac{m_\pi^2}{M_{n}^2}$ regardless of spin. But when the momentum is not significantly higher than excited states---for example, around $1$~GeV---the correlator formed from $(\bar{u}\gamma_t\gamma_5 d)$ receives more contamination $\propto \frac{m_\pi^2 E_{n,0}^2}{E_\pi^2M_{n,0}^2}$ from spin-$0$ states, and that formed from $(\bar{u}\gamma_z\gamma_5 d)$ receives more contamination $\propto \frac{m_\pi^2 E_{n,1}^2}{P_z^2M_{n,1}^2}$ from spin-$1$ states.

\begin{figure}[t]
	\centering
	\includegraphics[width=0.99\linewidth]{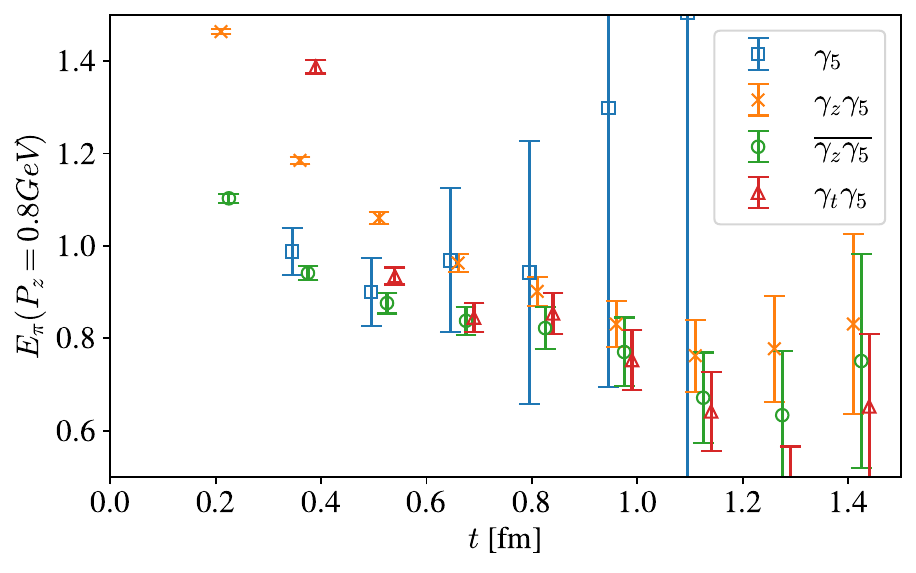}
	\caption{The effective mass plot of the pion two-point correlators at a boost of $0.8$~GeV using various interpolators.  At this moderate momentum, there is still obvious excited-state contamination from the axial-vector states in $(\bar{u}\gamma_z\gamma_5 d)$. However, the subtracted correlator, denoted by $\overline{\gamma_z\gamma_5}$, gives a good enhancement in the signal-to-noise ratio and also well-controlled excited-state contamination.} 
	\label{fig:spin1_subtraction}
\end{figure}

If we subtract the transverse counterpart in the correlator, we obtain
\begin{align}
	\langle [\bar{u}\gamma_z\gamma_5 d]^\dagger& [\bar{u}\gamma_z\gamma_5 d]\rangle-\langle [\bar{u}\gamma_x\gamma_5 d]^\dagger [\bar{u}\gamma_x\gamma_5 d]\rangle \nonumber\\
	&=\sum_{n,S=1} \frac{P_z^2}{M_{n,1}^2}C_{n,1} 
	+\sum_{n,S=0} \frac{P_z^2}{M_{n,0}^2}C_{n,0},
\end{align}
which remains positive definite (up to lattice artifacts, finite-volume effects, and statistical fluctuations), and also has a more suppressed excited-state contamination $\propto \frac{m_\pi^2 }{ M_{n}^2}$ at all momenta regardless of the spin of excited states. This suppression can reach $\mathcal{O}(10^{-2})$ for a physical pion, thus allowing an efficient extraction of the ground-state information, at the price of increasing the error by a factor of $\sqrt{2}$ that can be compensated by the large kinematic enhancement.

We test on a data set with both $(\bar{u}\gamma_z\gamma_5 d)$ and $(\bar{u}\gamma_x\gamma_5 d)$ interpolators and find excellent elimination of the axial-vector excited state, as shown in Fig.~\ref{fig:spin1_subtraction}. The  $\overline{\gamma_z\gamma_5}$ interpolator, defined as $(\bar{u}\gamma_z\gamma_5 d)$ after subtracting the transverse component, gives rise to correlators that show both significantly improved SNR and more suppressed excited state contamination. It may be useful in the study of heavy meson decays or $\pi\pi$ scattering, where the pion momentum is usually below $1$~GeV.

\section{Wick contraction contact terms}\label{sec:t=0}

Pion correlators for interpolators with generic Dirac structures $(\overline{u} \Gamma d)$ take the form
\begin{equation}
	C_{\pi}(\vec{P},t) = \sum_{\vec{x}} e^{\mathbf{i}\vec{P}\cdot\vec{x}}  \left< \overline{u}(t,\vec{x}) \Gamma d(t,\vec{x}) \overline{d}(0) \overline{\Gamma} u(0) \right>,
\end{equation}
where $\overline{\Gamma} \equiv \gamma_4 \Gamma^\dagger \gamma_4$ and in this section $\left< \cdot \right>$ denotes an expectation over fermionic degrees of freedom in some fixed gauge-field configuration. Applying Wick's theorem to this time-ordered products of fields gives
\begin{equation}
	\begin{split}
		C_{\pi}(\vec{P},t) &= -\sum_{\vec{x}} e^{\mathbf{i}\vec{P}\cdot\vec{x}} \text{Tr}\left[ S_u(\vec{x},t;0) \Gamma S_d(\vec{x},t;0) \overline{\Gamma} \vphantom{\frac{1}{2}}  \right. \\
		&\hspace{35pt} -  \left<  \mathcal{N}[d(\vec{x},t),\overline{d}(0)] \right> \overline{\Gamma}  S_u(\vec{x},t;0) \Gamma  \\
		&\hspace{35pt} + \left. \left< \mathcal{N}[\overline{u}(\vec{x},t), u(0)] \right> \Gamma S_d(\vec{x},t;0) \overline{\Gamma} \vphantom{\frac{1}{2}} \right],
	\end{split}
\end{equation}
where $\mathcal{N}$ denotes any finite-temperature definition of ``normal ordering'' as described in Ref.~\cite{Evans:1996bha}.
These expectation values of normal-ordered terms vanish for any fields with distinct spacetime arguments~\cite{Evans:1996bha}, giving
\begin{equation}\label{eq:meson_corr}
	\begin{split}
		C_{\pi}(\vec{P},t) &= -\sum_{\vec{x}} e^{\mathbf{i}\vec{P}\cdot\vec{x}} \text{Tr} \left[ S_u(\vec{x},t;0) \Gamma S_d(\vec{x},t;0) \overline{\Gamma} \right. \\
		&\hspace{30pt} + \delta_{0t} \left<  \mathcal{N}[\overline{u}(0), u(0)] \right> \Gamma S_d(0;0) \overline{\Gamma}  \\
		&\hspace{30pt} \left. - \delta_{0t} \left< \mathcal{N}[d(0),\overline{d}(0)] \right> \overline{\Gamma}  S_u(0;0) \Gamma \right].
	\end{split}
\end{equation}
At zero temperature, the normal-ordered expectation values become vacuum expectation values of fermion anticommutators, which can be explicitly evaluated using Eq.~(12) of Ref.~\cite{Luscher:1976ms}.
Antisymmetry of these anticommutators causes the two normal-ordered terms to add constructively in Eq.~\eqref{eq:meson_corr} and provide equal contributions in the isospin limit.
At non-zero temperature, they are more complicated to evaluate.

Without explicitly evaluating and including the normal-ordered terms in Eq.~\eqref{eq:meson_corr}, $C_{\pi}(\vec{P},t)$ can be identified with the usual expression $-\sum_{\vec{x}} e^{\mathbf{i}\vec{P}\cdot\vec{x}} \text{Tr}[S_u(\vec{x},t;0) S_d(\vec{x},t;0)]$ if and only if $t > 0$.
Note that this result only assumes that the valence quark fields are fermionic operators and holds for arbitrary discretization choices, even mixed/smeared actions, and the fact that $C_{\pi}(\vec{P}, 0) \neq -\sum_{\vec{x}} e^{\mathbf{i}\vec{P}\cdot\vec{x}} \text{Tr}[S_u(\vec{x},0;0) S_d(\vec{x},0;0)]$ is distinct from other concerns about ``contact terms'' arising from lattice-scale nonlocality in the action.
These  terms have significant effects on the hadron correlators made of kinematically-enhanced interpolators and without including them the spectral representations are badly violated and the Lanczos algorithm breaks down. 
A straightforward workaround is to include one application of the transfer matrix in the Lanczos initial state, i.e., start the analysis at $t=2a$.

\end{document}